\documentclass[prl,aps,twocolumn,showpacs,superscriptaddress]{revtex4}

\usepackage{graphicx}

\begin{document}

  \title{Spin-Orbit Coupling and Ion Displacements in     Multiferroic TbMnO$_3$}
  
  \author{H. J. Xiang}
  \affiliation{National Renewable Energy Laboratory, Golden, Colorado 80401, USA}
  
  \author{Su-Huai Wei}
  \affiliation{National Renewable Energy Laboratory, Golden, Colorado 80401, USA}

  \author{M.-H. Whangbo}
  \affiliation{Department of Chemistry, North Carolina State
    University, Raleigh, North Carolina 27695-8204, USA}

  \author{Juarez L. F. Da Silva}
  \affiliation{National Renewable Energy Laboratory, Golden, Colorado 80401, USA}

  \date{\today}

  \begin{abstract}
    The magnetic and ferroelectric (FE) properties of TbMnO$_3$ 
    were investigated on the basis of relativistic
    density functional theory (DFT) calculations.  
    We show that, due to spin-orbit coupling, the spin-spiral plane of TbMnO$_3$ 
    can be either the $bc$- or $ab$-plane, but not the $ac$-plane. 
    As for the mechanism of FE polarization, our work reveals that the ``pure electronic'' 
    model by Katsura, Nagaosa and Balatsky
    is inadequate in predicting the absolute direction of FE polarization.  
    Our work indicates that to determine the magnitude and the
    absolute direction of FE polarization in spin-spiral states,
    it is crucial to consider the displacements of the ions from their centrosymmetric
    positions.
    \end{abstract}


  \maketitle
  Recent studies on magnetic 
  ferroelectric (FE) 
  materials have shown that electric polarization 
  can be significantly modified by the application of a magnetic field
  \cite{Cheong2007,Kimura2003,Hur2004,Lawes2005,Katsura2005,Sergienko2006,Mostovoy2006}.
  Perovskite TbMnO$_3$ with a spin-spiral magnetic order is a prototypical multiferroic 
  compound with a gigantic magnetoelectric 
  effect \cite{Kimura2003}. 
  Currently, there are two important issues concerning the FE polarization of
  TbMnO$_3$.
  One concerns the origin of FE polarization. Model Hamiltonians studies of 
  spin-spiral multiferroic compounds have provided two different
  pictures.
  In the Katsura-Nagaosa-Balatsky (KNB) model \cite{Katsura2005}, 
  the hybridization of electronic states 
  induced by spin-orbit coupling (SOC) leads to a FE polarization of
  the charge density distribution even if the  
  ions are not displaced from their centrosymmetric positions. 
  In contrast, the model study by Sergienko and Dagotto
  \cite{Sergienko2006} concluded that oxygen ion displacements
  from their centrosymmetric positions 
  are essential for the FE polarization in multiferroic compounds \cite{Picozzi2007}. 
  When carried out with the ions kept at their centrosymmetric 
  positions, 
  density functional theory (DFT) 
  calculations \cite{Xiang2007}
  for the spin-spiral states of LiCuVO$_4$ predict FE polarizations that 
  agree reasonably well in magnitude
  with experiment \cite{Schrettle},
  which is in apparent support of the KNB model \cite{Katsura2005}. 
  It is, therefore, important to check which model, the KNB or the ``ion displacement'' model, 
  is relevant for the FE polarization in TbMnO$_3$.
  The other issue concerns the spin-spiral plane of TbMnO$_3$. Under a magnetic field, 
  the spin-spiral plane of TbMnO$_3$ can be either the
  $bc$-plane or the $ab$-plane, but not the $ac$-plane. 
  To explain this observation, it is necessary to probe 
  the magnetic anisotropy of the Mn$^{3+}$ ion. 
  The magnetic anisotropy of the Tb$^{3+}$ ion
  might
  be also relevant for the
  magnetoelectric effect, as suggested by Prokhnenko {\it et al.}
  \cite{Prokhnenko2007}.

  In this Letter, we investigated these issues
  on the basis of
  DFT calculations and found that the consideration of the ion
  displacements is essential
  for the FE polarizations in the spin-spiral state of TbMnO$_3$, and
  the KNB model can be erroneous even for predicting the absolute
  direction of FE polarization. The absence of the $ac$-plane spin-spiral in
  TbMnO$_3$ is explained by the magnetic anisotropy of the Mn$^{3+}$ ion.

\begin{figure}
    \includegraphics[width=4.5cm]{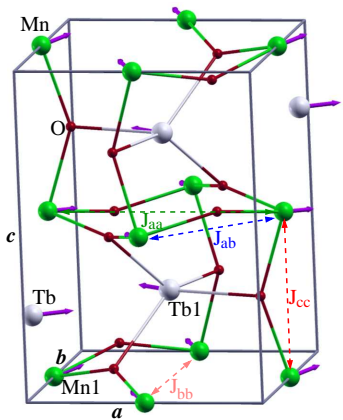}
    \caption{ (Color online)
      Perspective view of the orthorhombic structure of TbMnO$_3$.
      The large, 
      medium, and small spheres represent the Tb, Mn, and O
      ions, respectively.
      The Mn-Mn spin exchange paths $J_{ab}$, $J_{aa}$, $J_{bb}$, and
      $J_{cc}$ are also indicated. The solid vectors denote the 
      easy axes for 
      the Tb$^{3+}$ and Mn$^{3+}$ spins.
      }
    \label{fig1}
  \end{figure}

  Our calculations were based on DFT plus
  the on-site 
  repulsion U method \cite{Liechtenstein1995} within the generalized gradient
  approximation  \cite{Perdew1996}. 
  We used $U_{eff}=6.0$ eV on Tb 4f states. With other $U_{eff}$ values for Tb,
  similar results were obtained. 
  For Mn 3d states,  we employed $U_{eff}=2.0$ eV, which leads to the 
  spin exchange interactions between the Mn$^{3+}$ ions that are 
  consistent with the observed magnetic structure of TbMnO$_3$ (see below).
  For the calculation of FE polarization, the Berry phase method 	
  \cite{King-Smith1993} encoded in
  the Vienna ab initio simulation package (VASP) was
  employed \cite{VASP,PAW}, 
  in which the Tb 4f electrons were
  treated as core electrons. For the study of the SOC
  effect associated with 
  the Tb 4f electrons, we used the
  full-potential augmented plane wave plus local orbital method as
  implemented in the WIEN2k code \cite{wien2k}. 
  Due to the small value of the spin anisotropy, 
  we employed the convergence threshold of $10^{-7}$ for electron density. 
  As shown in Fig.~\ref{fig1}, the experimental 
  crystal structure \cite{Alonso2000} of
  TbMnO$_3$ has a distorted 
  GdFeO$_3$-type orthorhombic perovskite structure with space group Pbnm. 
  In our calculations, the experimental structure 
  was used unless otherwise stated.

  \begin{figure}
    \includegraphics[width=6.5cm]{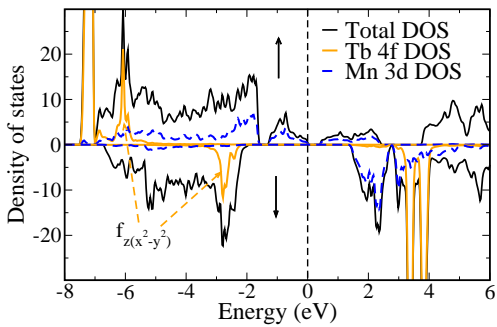}
    \caption{(Color online) Electronic structure 
      obtained for the FM
      state of TbMnO$_3$ from the WIEN2k 
      calculations. 
      The total DOS, the Tb 4f partial DOS, and the Mn 3d partial DOS are shown. The position
      of the $f_{z(x^2-y^2)}$ state is also indicated.}
    \label{fig2}
  \end{figure}

  The basic electronic structure of TbMnO$_3$
  is shown in Fig.~\ref{fig2} 
  in terms of 
  the density of states (DOS) of the FM state calculated 
  by WIEN2k with the Tb 4f states treated as valence states.
  The system is an insulator with 
  energy gap of about 0.5 eV between the 
  spin-up 
  Mn 3d $e_g$ states.
  The spin-up 
  Tb 4f states are all occupied while
  the spin-down Tb 4f states
  occur at around $-2.8$ eV with the remaining down-spin 
  Tb 4f states above the Fermi level.  
  This feature is consistent with the f$^{8}$ configuration of the Tb$^{3+}$ ion. 
  Partial DOS analysis indicates that the occupied 
  down-spin states have mainly the $f_{z(x^2-y^2)}$ orbital character with a 
  slight contribution from the $f_{xyz}$   orbital.
  
  For the spin exchange between Mn$^{3+}$ ions, there are four spin exchange paths to 
  consider as shown in Fig.~\ref{fig1}:
  $J_{ab}$ is the intralayer ($ab$ plane) nearest neighbor (NN) exchange interaction,
  $J_{aa}$ and
  $J_{bb}$ are the intralayer next-nearest neighbor (NNN) exchange
  interactions along the $a$ and $b$ directions, respectively, and $J_{cc}$ 
  is the interlayer NN exchange interaction along the $c$ direction. The high-spin
  Mn$^{3+}$ 
  ions have the $t_{2g}^3 e_g^1$ configuration with the $e_g$ states forming a
  staggered
  orbital ordering of the $d_{3x^2-r^2}$ and $d_{3y^2-r^2}$ states. 
  Such an orbital ordering induces an NN intralayer ferromagnetic (FM) exchange, 
  and an NN interlayer antiferromagnetic (AFM) exchange.
  The cooperative Jahn-Teller distortion 
  associated with the orbital ordering leads to a large NNN intralayer 
  super-superexchange 
  along the $b$ direction \cite{Kimura2003B,Picozzi2006}. Our VASP calculations 
  lead to $J_{ab}=-1.52$ meV, $J_{aa}=0.57$ meV, $J_{bb}=0.85$ meV, and
  $J_{cc}=0.50$ meV. 
  The calculated $J_{ab}$, $J_{bb}$ and $J_{cc}$ values are 
  consistent with those expected 
  for TbMnO$_3$ \cite{Kimura2003B}. 
  $J_{aa}$ was predicted to be FM by 
  Kimura {\it et al.} \cite{Kimura2003B}, 
  but our calculations show it to be AFM.
  The classical spin analysis based on the Freiser method
  \cite{Freiser1961} predicts
  that the spin ground state is
  a spin spiral with the modulation vector $\mathbf{q}=(0,q_y,0)$ with
  $q_y=\mathrm{acos}(-\frac{J_{ab}}{2 J_{bb}})/\pi=0.15$.
  The prediction of a spin-spiral ground state is in agreement with 
  the experimental observation \cite{Kenzelmann2005}, which shows 
  an incommensurate spiral with $\mathbf{q}=(0,0.27,0)$ below 28 K.
  It is worthwhile to point out that if a smaller $U_{eff}$ is used
 $J_{bb}$ will be stronger, and $J_{ab}$ be weaker, hence leading to
  a larger $q_y$.

\begin{figure}
    \includegraphics[width=6.5cm]{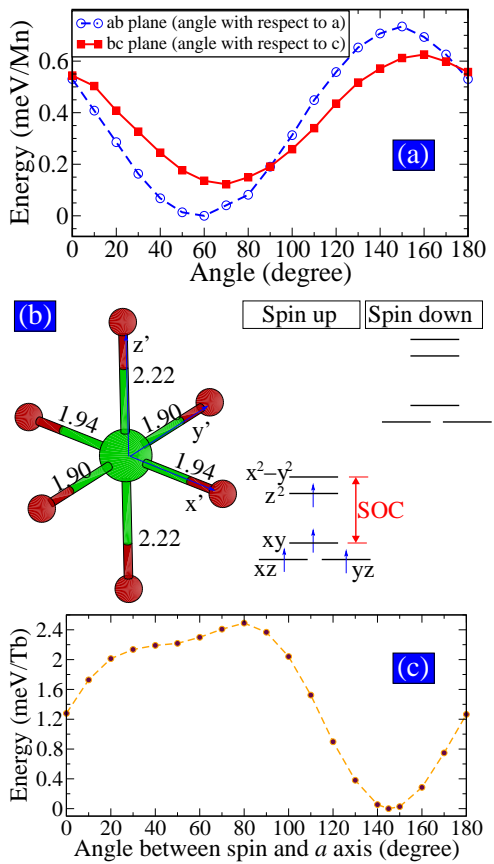}
    \caption{(Color online) (a) The 
    dependence of energy on the direction of the Mn spin
    in the $ab$-plane or the $bc$-plane. 
    (b) The local structure
    of the distorted MnO$_6$ octahedron. The numbers give the Mn-O
    bond lengths in \AA. The local coordinate system is also
    indicated. The right-hand-side diagram illustrates the 
    electron configuration of the Mn$^{3+}$ ($d^4$) ion. 
    The label
    ``SOC'' denotes that the largest SOC
    mixing
    occurs between the $d_{xy}$ and $d_{x^2-y^2}$
    states. (c) The 
    dependence of energy on the direction of Tb spins 
    in the $ab$-plane.}
    \label{fig3}
  \end{figure}

  To examine the magnetic anisotropy of the high-spin Mn$^{3+}$ ion in TbMnO$_3$, 
  it is necessary to consider DFT+U calculations with SOC  included. 
  TbMnO$_3$ has four Mn$^{3+}$ ions
  per unit cell. The easy axes of these ions are not collinear 
  but related by symmetry. 
  Since the effect of SOC is largely local in nature,
  we consider for simplicity 
  only one Mn ion (i.e., Mn1 in Fig.~\ref{fig1}) per unit cell
  by replacing the remaining three Mn$^{3+}$ ions  
  with Sc$^{3+}$ ions
  that have no magnetic moment. To remove 
  any possible coupling with the Tb 4f moment, we replace 
  the Tb$^{3+}$ ions of the unit cell with La$^{3+}$ ions 
  that have no magnetic moment. 
  The energy dependence upon the Mn spin direction obtained from DFT+U+SOC calculations 
  is shown in Fig.~\ref{fig3}(a) for the cases when the Mn spin 
  lies in the $ab$ and $bc$ planes.
  The energy minimum occurs at 60$^\circ$ and 70$^\circ$ for 
  the case when the spin lies 
  in the $ab$ and $bc$ planes, respectively, and the energy minimum 
  for the $ab$-plane is lower than that for $bc$-plane by $0.12$ meV/Mn. 
  
  The above results can be readily understood by
  analyzing the effect of the SOC Hamiltonian \cite{Xiang2007B,Xiang2008}. As
  shown in Fig.~\ref{fig3}(b), each MnO$_6$ octahedron of TbMnO$_3$ is axially elongated. 
  With one of the two longest Mn-O bonds 
  taken as the local $z$ axis and neglecting the 
  slight difference between the other four short Mn-O bonds, the d-block levels of
  the Mn$^{3+}$ (d$^4$) ion are described as shown on the
  right 
  hand side of Fig.~\ref{fig3}(b), which leads to the electron
  configuration 
  $(xz)^{1}(yz)^{1}(xy)^{1}(z^2)^{1}(x^2-y^2)^{0}$. 
  The SOC Hamiltonian $\lambda \hat{\mathbf {L}} \cdot \hat{\mathbf
    {S}}$ leads to 
  an interaction of the empty
  $d_{x^2-y^2}$ ($e_g$) state 
  with the other d-states.
  The strongest SOC occurs between two $e_g$ states ($d_{xy}$ and
  $d_{x^2-y^2}$) with the maximum energy lowering when the spin 
  lies in the local $z$ direction \cite{Xiang2008}. For the Mn1 ion, the $\theta$ and 
  $\phi$ angles of the local $z$ direction in the global coordinate system are
  $\theta=80.2^\circ$ and $\phi=60.5^\circ$, 
  respectively. 
  In good agreement with these values, our 
  SOC analysis based on tight-binding calculations 
  \cite {Harrison} shows that the actual easy axis
  of Mn1 has $\theta=84^\circ$ and $\phi=60^\circ$.
  Thus the easy axis is
  close to the $ab$ plane, and is far from the $c$ axis. This 
  explains why the energy minimum for the Mn spin lying in the $ab$-plane has 
  a lower energy than that lying in the $bc$-plane. 
  In addition, the easy axis 
  is closer to the $b$ axis than 
  to the $a$ axis. Our result is consistent with the experimental
  observation \cite{Quezel1977} that 
  the sine-wave modulation of the Mn magnetic moment below 40 K 
  has the direction parallel to the $b$-axis. 
  Since the easy axis of the Mn spin is far from the $c$ direction,
  it is expected that the spin-spiral plane of 
  TbMnO$_3$ is either the $ab$-plane or the $bc$-plane, but not the
  $ac$-plane \cite{test}.  Experimentally \cite{Kenzelmann2005}, it
  was found that 
  the Mn moments of TbMnO$_3$ 
  form a $bc$-plane elliptical spiral with $m_b$(Mn) $= 3.9$ $\mu_B$ $>$
  $m_c$(Mn) $= 2.8$ $\mu_B$. This can be readily explained by 
  the fact that
  the easy axis of the Mn spins is close to the $b$ axis, but
  far from the $c$ direction. 

  \begin{figure}
    \includegraphics[width=8.0cm]{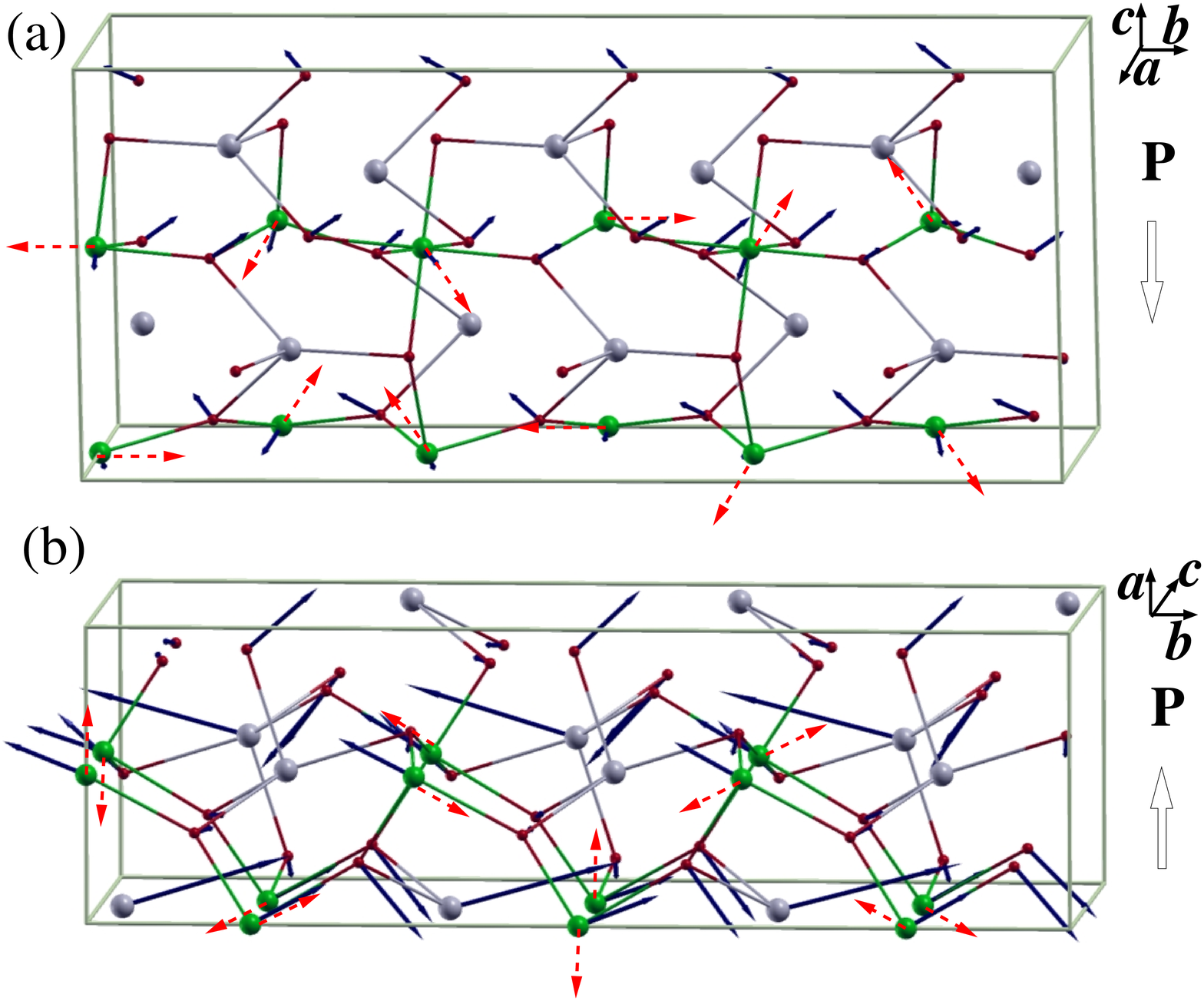}
    \caption{(Color online) The atomic displacements (indicated by
      solid arrows)
      of the spin-spiral states after 
      geometry optimization 
      for (a) the  $bc$-plane spiral and (b) the $ab$-plane spiral.
      The 
      dashed arrows
      indicate the directions of the Mn spins 
      in the spiral states. The
      directions of the electric polarizations are also shown.}
    \label{fig4}
  \end{figure}

  The magnetic anisotropy of the 
  Tb$^{3+}$ ion in TbMnO$_3$ was also calculated in a similar manner;
  three of the four Tb$^{3+}$ ions in a unit cell were replaced by La$^{3+}$ ions 
  with all the Mn$^{3+}$ ions of a unit cell replaced by Sc$^{3+}$ ions. 
  Our DFT+U+SOC calculations show that the Tb spin prefers 
  to lie in the $ab$ plane since the 
  state for the spin parallel to the 
  $c$-axis is higher 
  than that for the spin in the $ab$ plane by at least 0.83 meV/Tb. The energy calculated 
  for Tb1 (as labeled in Fig.~\ref{fig1} as a function of the angle $\phi$ with 
  $\theta =90^\circ$ is presented in Fig.~\ref{fig3}(c), 
  which reveals that the energy miminum occurs at
  around $\phi=145^\circ$. 
  The large anisotropy energy is consistent with the Ising behavior of
  Tb moment \cite{Prokhnenko2007}. The easy axes of the Tb$^{3+}$ ions 
  presented in Fig.~\ref{fig1} show that they lie symmetrically
  around the $b$-axis with the angle $55^\circ$.
  This is in excellent agreement with the observed angle of $57^\circ$ 
  \cite{Quezel1977}.

  The electric polarization of TbMnO$_3$ in the 
  spin-spiral state was calculated using 
  the VASP. 
  To reduce the computational 
  task, we considered the $\mathbf{q}=(0,1/3,0)$ state, 
  which we simulated by using a $1\times 3 \times 1$
  supercell.  
  Using the experimental centrosymmetric
  structure, the electric polarization from the pure 
  electronic effect
  was calculated to be $\mathbf{P}=
  (0,0,0.5)
  $ $\mu C/m^2$
  \cite{polarization, Seki2008, Malashevich2008} 
  for the $bc$-plane spiral shown
  in Fig.~\ref{fig4}(a). Experimentally, the magnitude of the electric polarization
  for the $bc$-plane spiral is about 600 $\mu C/m^2$
  \cite{Prokhnenko2007B}, 
  which is three orders of
  magnitude larger than the value calculated with no geometry relaxation.
  For the $ab$-plane spiral shown in Fig.~\ref{fig4}(b), 
  the electric polarization is calculated to be 
  $\mathbf{P}=
  (-331.0,0,0)
  $ $\mu C/m^2$
  in the absence of geometry relaxation. 
  The absolute directions of the FE polarizations obtained for the $bc$- and $ab$-plane 
  spin-spiral states from DFT calculations without geometry relaxation are
  opposite to those predicted by the KNB model (Eq. 6 of ref. \cite{Katsura2005}). 
  To examine the effect of ion displacements in the spin-spiral states 
  on the FE polarization, we optimized the atom positions of TbMnO$_3$ in the
  $bc$- and $ab$-plane spiral states by performing DFT+U+SOC calculations and 
  calculated the electric polarizations of TbMnO$_3$ using the relaxed structures. 
  These calculations lead to $P_c$ = $
  -424.0
  $ $\mu C/m^2$ for the $bc$-plane
  spiral, and to $P_a$ = $
  131.2
  $ $\mu C/m^2$ for the $ab$-plane spiral. 
  The absolute directions of the FE polarizations are switched for both
  spin-spiral states under geometry relaxation.
  The calculated FE polarization for
  the $bc$-plane spiral using the relaxed
  structure is now much closer to the experimental value.
  Furthermore, the direction of the FE polarization is
  in agreement with experiment 
  \cite{Yamasaki2007B, polarization, Seki2008, Malashevich2008} 
  as well as the KNB model. However, 
  this agreement between the KNB model and experimental values is fortuitous;
  for LiCuVO$_4$ and LiCu$_2$O$_2$, which consists of CuO$_2$ ribbon
  chains, the KNB model predicts  
  the wrong absolute direction of the FE
  polarization for the 
  $ab$-plane spiral, 
  but the correct absolute
  direction of the FE polarization for the 
  $bc$-plane spiral, 
  regardless of whether the
  unrelaxed or the relaxed crystal structures are employed
  \cite{Xiang2007}.
  The failure of the KNB model could be due to the 
  fact that it was derived for the
  $t_{2g}$ 
  systems while the $e_g$ states 
  are important in Mn$^{3+}$ and
  Cu$^{2+}$ systems. 
  Furthermore, we notice that the FE polarization of the $ab$-plane spiral
  calculated by using the relaxed structure, 
  now along the 
  $a$ 
  direction, 
  is smaller in magnitude than 
  that calculated by using the experimental
  structure. 
  In the case of LiCuVO$_4$ and LiCu$_2$O$_2$, the geometry relaxation in 
  the spin-spiral states was found to enhance the magnitudes of the FE polarizations
  without changing their absolute directions \cite{Xiang2007}. 
  
  To show how the ion displacements break the centrosymmetry, we
  present the pattern of the atom 
  displacements with respect to the centrosymmetric structure in
  Fig.~\ref{fig4}. For the $bc$-plane 
  spiral, the O atoms at the Wyckoff 4c position (i.e., O1) lying in the 
  $ab$ plane of the Tb atoms almost do not move. The movement of all the O atoms
  at the Wyckoff 8d position (i.e., O2) close to the $ab$ plane of the Mn atoms
  have a component along the $c$ direction.
  All the Mn and Tb atoms have a displacement along the $-c$
  direction. The largest 
  displacements along the $c$ direction (about $2.4\times
  10^{-4}$ \AA) occur 
  at the Mn sites rather than at the O sites. This finding is 
  in contradiction to
  the assumption introduced in the ``ion-displacement'' model 
  \cite{Sergienko2006}. Considering that the Born effective 
  charge is positive for 
  the Tb$^{3+}$ and Mn$^{3+}$ cations, and negative for the O$^{2-}$ anion,
  the total electric polarization is 
  expected to be along the 
  $-c$ direction, in agreement with the DFT calculation. For the
  $ab$-plane spiral, 
  some O atoms have displacements 
  along the $-a$ direction, but other O atoms have displacements along the $a$
  direction. The sum of the O ion displacements is along the $-a$
  direction. The occurrence of  
  alternating O displacements 
  is consistent with
  the prediction made by Sergienko and Dagotto \cite{Sergienko2006},
  and is responsible for the smaller electric polarization when
  compared with the case of the $bc$-plane spiral. All the Mn atoms have a
  displacement along the $a$ direction. Another unexpected finding is
  that some Tb atoms have the largest displacements with a large component
  along the $b$ or $-b$ direction and a small component along the $a$
  direction. 
     
  In summary, 
  the absence of the $ac$-plane spin-spiral in 
  TbMnO$_3$ is explained 
  by the magnetic anisotropy of the Mn$^{3+}$ ion. The calculated easy
  axis for Tb is in excellent agreement with the experimental result. 
  The consideration of the ion displacements in the
  spin-spiral states of TbMnO$_3$ is
  essential in determining the magnitude and 
  the absolute direction of the FE
  polarizations, which is in support of the ``ion-displacement''  model. 
  Surprisingly, 
  the displacements of the Mn$^{3+}$ and Tb$^{3+}$ ions are
  generally greater than those of the O$^{2-}$ ions.  
  The KNB model, however, can fail to describe both the magnitude and the absolute direction of 
  FE polarization.

  Work at NREL was supported by the U.S. Department of
  Energy, under Contract No. DE-AC36-99GO10337.
  The research at NCSU was supported by the Office of Basic
  Energy Sciences, Division of Materials Sciences, U.S.
  Department of Energy, under Grant No. DE-FG02-86ER45259. 
  We thank Prof. D. Vanderbilt for the discussion about the definition of
  positive electric polarization
  \cite{Yamasaki2007B, Seki2008, Malashevich2008}.




\begin{thebibliography}{99}
\bibitem{test}VASP 
    calculations show that the $ac$-plane spiral has
    higher energy by 0.23 meV/Mn than the $ab$-plane or $bc$-plane
    spiral. The energy difference between the $ab$-plane spiral and
    $bc$-plane spiral is within 0.02 meV/Mn. This difference is much
    smaller than that (0.12 meV/Mn) between the
    minima on the $bc$ and $ab$ plane since in a spiral not all
    spins lie along the minimum direction.
\bibitem{polarization}In this work 
    we adopt the conventional definition of the 
    positive direction of electric polarization (i.e.,     
    from the negative to the positive charge site).
\end{thebibliography}
\end{document}